\documentclass[10pt,twocolumn,graphicx]{article}
\usepackage{graphicx}

\date{}
\begin{document}

\title{{\bf Quantum-Implemented Selective Reconstruction
of High-Resolution Images}}

\author{Mitja Peru\v s \ \& \ Horst Bischof  \\
        Institute for Computer Vision \& Graphics,
        Graz University of Technology \\
        Inffeldgasse 16 (2. OG), A-8010 Graz, Austria \ / \
        \{perus,bischof\}@icg.tu-graz.ac.at
        \and Loo Chu Kiong \\
        Faculty of Engineering and Technology,
        Multimedia University (Melaka) \\
        MY-75450 Melaka, Malaysia \ / \
        ckloo@mmu.edu.my
        }

\maketitle \thispagestyle{empty}

\begin{abstract}
{\it This paper proposes quantum image reconstruction.
Input-triggered selection of an image among many stored ones, and
its reconstruction if the input is occluded or noisy, has been
simulated by a computer program implementable in a real
quantum-physical system. It is based on the Hopfield associative
net; the quantum-wave implementation bases on holography. The main
limitations of the classical Hopfield net are much reduced with
the new, original -- quantum-optical -- implementation. Image
resolution can be almost arbitrarily increased.}
\end{abstract}

\section{Introduction}

There is growing evidence that quantum-physical systems could be
harnessed for information processing, including specifically image
recognition, in two ways:
\begin{itemize}
\setlength{\partopsep}{0mm} \setlength{\topsep}{0mm}
\setlength{\parskip}{0mm} \setlength{\parsep}{0mm}
\setlength{\itemsep}{0mm} \item by Turing-machine-based quantum
computing using quantum logic gates
\cite{trugenberger,ventura,ventopt,sch}; \item by quantum
processing similar to those in (oscillatory) associative neural
nets \cite{q} (cf. \cite{ron,hnet}).
\end{itemize}
This paper reports how it is possible to implement successful
image recognition, as verified by our simulations, in a quantum
holographic process \cite{qhol}. Since the natural fundamental
quantum-wave dynamics is harnessed, it allows much easier and
cheaper physical realization with much bigger sizes and
resolutions of images than the mainstream quantum-computing
approaches \cite{trugenberger,ventura,ventopt,sch}.

The main contribution of this paper is not to propose a {\it
generally}-better image-recognition method, but to present its
{\it powerful alternative implementation into a quantum-wave
medium} (sec. 2), and to demonstrate its plausibility by
computational experiments (sec. 3). Quantum-net's capacities of
connectivity, parallelism, storage, associativity, speed and
miniaturization are {\it enormous}, even much greater than in
classical holography \cite{hol,opr}.

In \cite{cary} it was shown how the Hopfield model with
real-valued (thus not necessarily binary) activities of units /
neurons, having linear (not sigmoid or signum) activation
function, can be transformed into a quantum-holographic procedure
\cite{qhol} where the Hebbian memory-storage is replaced by
multiple self-interferences of quantum plane-waves. This
translation succeeded by the simplest variable exchange of the
Hopfield's real-valued variables with the complex-valued variables
changing according to sinusoids (waves) (cf. \cite{ron,hnet}).
Thereby, {\it all input-to-output transformations are preserved}.
Thus, {\it quantum-wave image recognition functions equivalently
to Hopfield's one}, only the implementation is much miniaturized
enabling {\it almost infinitely large} neural-like networks.

Since the opposite translation, i.e. digitalization of holography,
was done in the sixties of the $20^{th}$ century to get the first
computational associative memories, one might wonder what is new
in the present proposal. The big experimental success of classical
(optical, acoustic, microwave- etc., but also X-ray-, atom-,
electron-) holography \cite{hol} is widely known, but not also the
recent fast development of quantum optics \cite{qopt} which gave
birth to quantum holography \cite{qhol} (good "tutorial" in
\cite{granik}). The latter promises to implement the well-known
Hopfield model and its generalizations in a completely new
framework where the former obstacles (memory-capacity limitations,
problems with non-orthogonality of small-size inputs producing
cross-talk) are very much reduced.

\section{Web of quantum waves}

Using neuro--quantum "isomorphisms", presented systematically in
\cite{ana}, and "numbers-to-waves" translation, as in \cite{cary},
we transform the Hopfield-like associative neural net into quantum
formalism (details in \cite{cary,q}): \begin{itemize}
\setlength{\partopsep}{0mm} \setlength{\topsep}{0mm}
\setlength{\parskip}{0mm} \setlength{\parsep}{0mm}
\setlength{\itemsep}{0mm}
\item Quantum wave-function $\Psi$ acts
as net's state vector $\vec{q}$. \item Eigen-wave-functions
$\psi^k$ ($k=1,...,P$) act as Hopfield's pattern-bearing
eigen-vectors (attractors) $\vec{v}^k$. \item The quantum
Green-function propagator {\bf G} replaces the Hebb memory matrix
{\bf J}. \item Thus, sum of self-interferences $\psi^k \otimes
\psi^k$ of quantum waves $\psi^k$ (that's the "hologram" {\bf G})
implements the sum of auto-correlations of input-pattern
configurations $ \vec{v}^k \otimes \vec{v}^k $ (that's the
content-addressable associative memory {\bf J}). ($\otimes$
denotes tensor/outer product.)
\end{itemize}
\setlength{\partopsep}{0mm} \setlength{\topsep}{0mm}
\setlength{\parskip}{0mm} \setlength{\parsep}{0mm}
\setlength{\itemsep}{0mm} The Hebb-equivalent expression for
elements of {\bf G} (i.e., the multiple cris-cross array $ \sum_k
\psi^k \otimes \psi^k $ implementing matrix {\bf J}) is:
$$ G_{hj} = \sum_{k=1}^P \psi_h^k (\psi_j^k)^{\ast}
\eqno(1)
$$ where $h$ and $j$ denote the unit / pixel / "neuron" /
quantum point at locations $\vec{r}_1$ and $\vec{r}_2$ at time $t$
($ h, j = 1,..., N$; $N$ can be almost infinite). The asterisk
denotes complex conjugation (or, optically, phase conjugation).

After we have succeeded to encode patterns or images as
eigen-states (attractors) $\psi^k$ into the quantum system
prescribed by eq. (1), we can reconstruct one (say, $k_0^{th}$) by
presenting a new input similar to the $k_0^{th}$ stored one:
$$ \Psi_h^{output} = \sum_{j=1}^N G_{hj} \Psi_j^{input}
 = \sum_{j=1}^N
\left( \sum_{k=1}^P \psi^k_h (\psi^k_j)^{\ast} \right)
\Psi_j^{input} = $$
$$ = \sum_{k=1}^P \left( \sum_{j=1}^N (\psi_j^k)^{\ast} \Psi_j^{input}
\right) \psi_h^k \ \ \doteq \ \ \psi_h^{k_0} \eqno(2)  $$
describes the resulting {\it selective retrieval (recognition) of
image} $\vec{v}^{k_0}$ encoded in $\psi^{k_0}$. See detailed
analysis in \cite{cary} or \cite{q}. Eq. (2) is in analogy with
\cite{vapnik}. In the quantum Dirac notation, eq. (2) is, using $
 (\vec{a} \otimes \vec{b}) \vec{c} = \langle \vec{b}, \vec{c}
\rangle \vec{a} $:
$$ \mid \Psi^{output} \rangle = {\bf G} \mid \Psi^{input} \rangle
= ( \sum_k \mid \psi^k \rangle \langle \psi^k \mid ) \mid
\Psi^{input} \rangle = $$ $$ = \sum_k \langle \psi^k \mid
\Psi^{input} \rangle \mid \psi^k \rangle \ \ \doteq \ \ \psi^{k_0}
\eqno(3) . $$

We assume that we can encode images $\vec{v}^k$ into {\it quantum
plane waves} (i.e., propagating sinusoidally-changing
probability-distribution for measuring a photon \footnote[1]{The
$k^{th}$ mode of the photon has momentum $\vec{p}^k$ and energy
$E^k$; $\hbar$ is Planck's constant; $i = \sqrt{-1}$} at location
$ \vec{r}$ at time $t$): $$ \psi^k (\vec{r}, t) = A^k (\vec{r}, t)
e^{i \varphi^k (\vec{r}, t)} = A^k e^{\frac{i}{\hbar}(\vec{p}^k
\vec{r} - E^k t)} \eqno(4) .
$$ We {\it may} choose the {\it same} constant amplitudes $A$, so
that quantum phases (delays between wave-peaks) $\varphi$ encode
the whole information. Let us take $A=1$ (or $A=1/\sqrt{N}$ for
convenient quantum normalization); so, $A^k_j = 1$ (or another
constant) for all $k, j$. (Various possibilities of amplitude and
phase modulation see in \cite{cary}.) The image-modulated
laser-beam is thus:
 $ \psi^k = (e^{i \varphi_1^k}, e^{i \varphi_2^k}, ..., e^{i
\varphi_N^k}) $ where the number of wave-front points (wave peaks)
is $N$.

The "isomorphism" of \cite{cary} allows us to exchange variables,
$ \vec{v}^k \leftrightarrow e^{i \varphi} $, giving $\psi^k_j =
e^{i \varphi^k_j}$ instead of Hopfield-like $\psi^k_j = v^k_j$ (or
$\psi_j^k = A_j^k$, respectively). With this exchange in equations
(1) and (2), all the information-processing mathematics, verified
by computer experiments of sec. 3 and \cite{simul}, remains valid
for sinusoid-encoded images also. Namely, because eq. (1) becomes
$$ G_{hj} = \sum_{k=1}^P e^{i \varphi_h^k} e^{-i \varphi_j^k} =
\sum_{k=1}^P e^{i (\varphi_h^k - \varphi_j^k)} \eqno(5) , $$ eq.
(2) becomes
$$ e^{i \varphi_h^{output}} =
\sum_{j=1}^N \left( \sum_{k=1}^P e^{i \varphi_h^k} e^{-i
\varphi^k_j} \right) e^{i \varphi_j^{input}} = $$ $$ =
\sum_{k=1}^P \left( \sum_{j=1}^N e^{-i \varphi_j^k} e^{i
\varphi_j^{input}} \right) e^{i \varphi^k_h} \ \ \doteq \ \ e^{i
\varphi_h^{k_0}} \eqno(6) .
$$

If images are almost orthogonal, a wave carrying an image (those
among many stored ones which is the most similar to the newly
input one) is {\it selectively reconstructed}.

There is a {\it non-local} information-exchange involved in this
holographic process, which in our quantum case exploits the
quantum interference web ({\bf G}) itself, not its static imprint
onto a crystal plate as in classical holography \cite{hol}.

Our information-processing result {\it can be extracted} from
$\psi^{k_0}$ using new quantum-optical (and computer-aided)
techniques for "ensemble"-measurement of observables or for
quantum-holographic-(like) wavefront reconstruction \cite{qwfe}:
\\
$\bullet$ quantum-phase estimation / engineering, \\ $\bullet$
wave-packet / wave-function \ reconstruction / sculpting / engineering, \\
$\bullet$ (coherent) quantum control / manipulation, \\ $\bullet$
quantum tomography.

\begin{figure*}[ht]
 \begin{tabular}[t]{cc}
   \centering
   \includegraphics[width=\columnwidth]{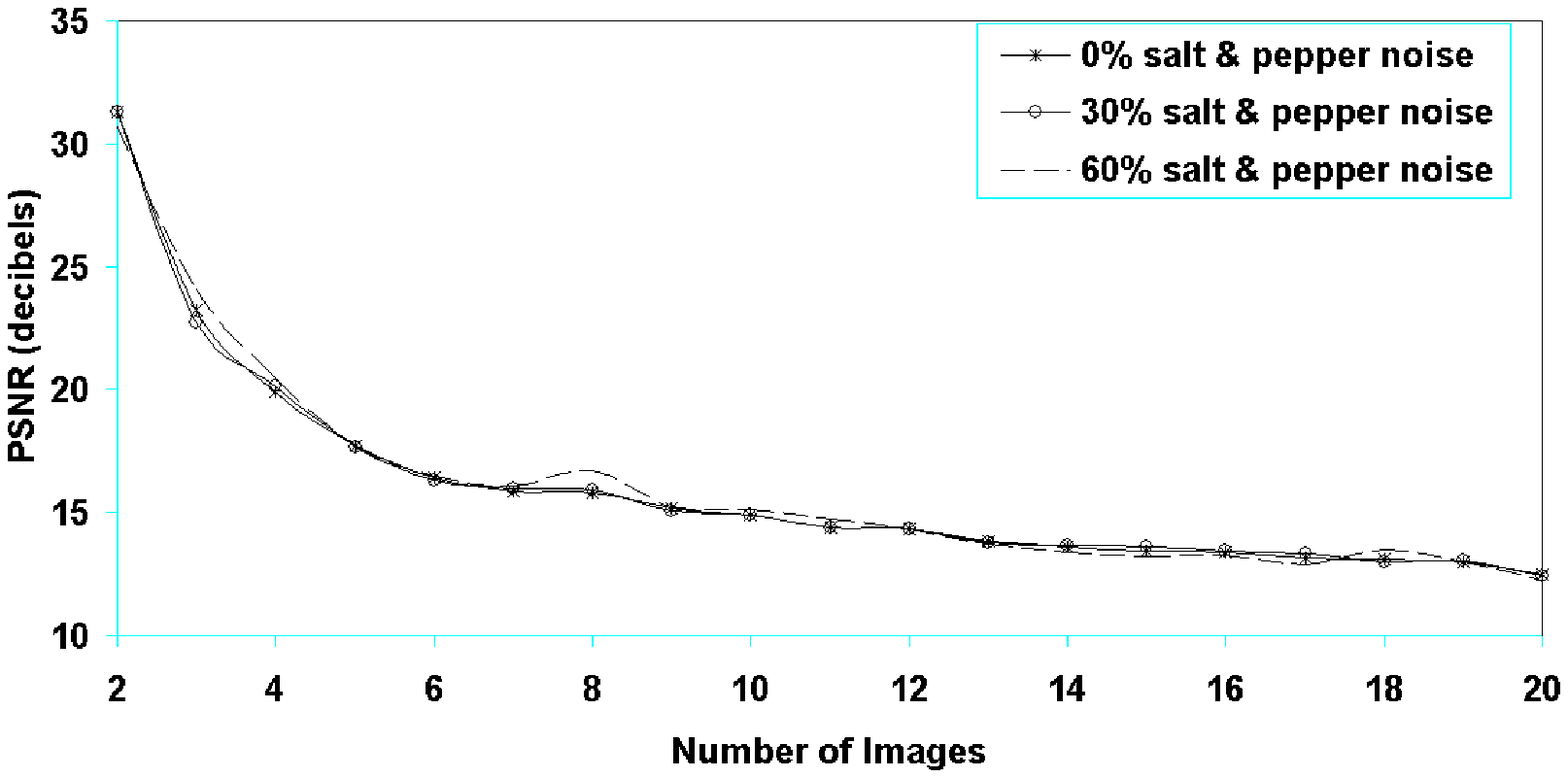} &
   \includegraphics[width=\columnwidth]{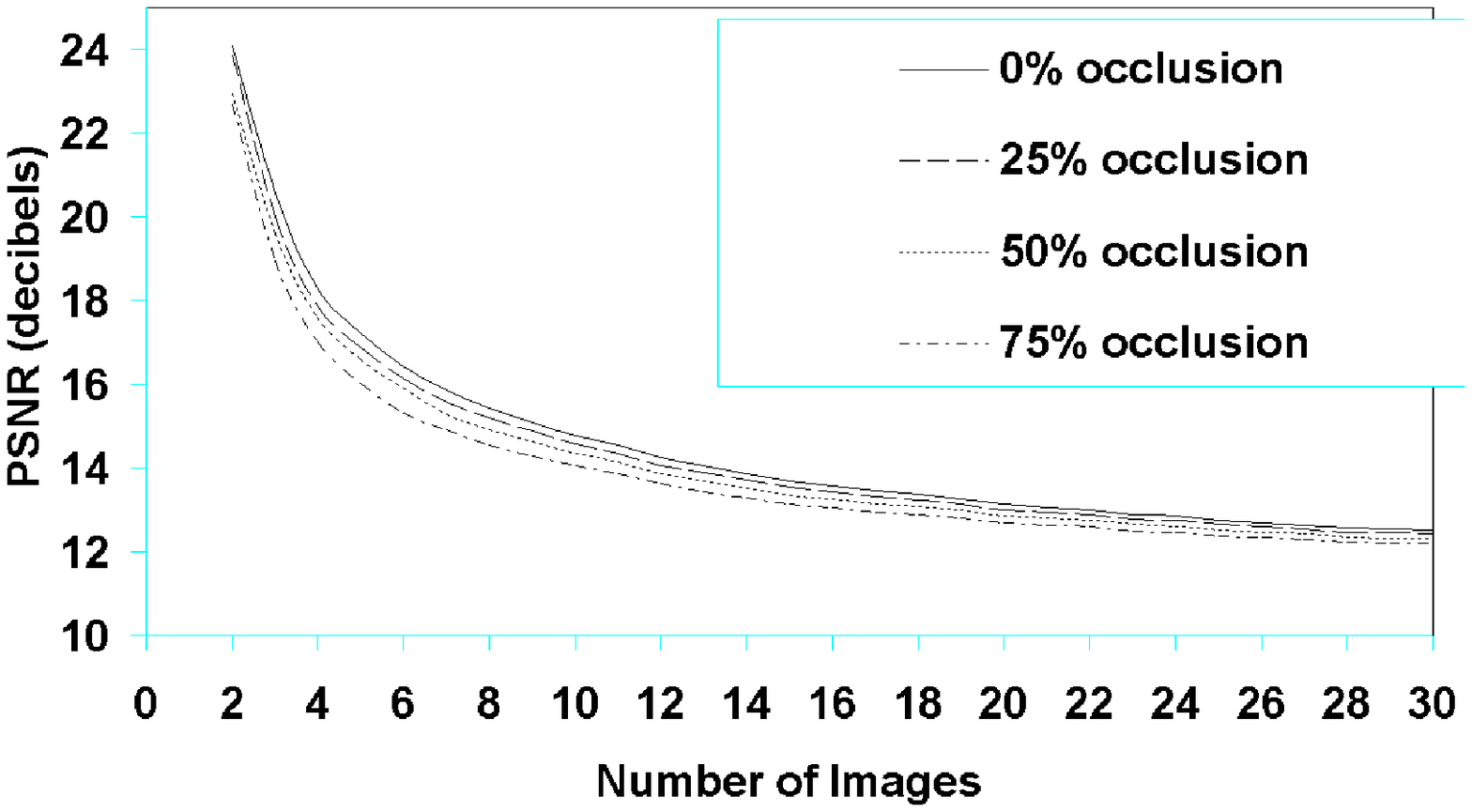} \\
 \end{tabular}
   \caption{Plots of {\bf Peak Signal-to-Noise Ratio of reconstructed
   image from "query-image" {\it versus} number of
   simultaneously-stored images} of {\bf (left)} Chinese pictograms and
   {\bf (right)} fingerprints, where {\bf (left)} query is a Chinese
   pictogram with salt-and-pepper noise, and {\bf (right)} query is
   an occluded fingerprint}
   \label{plots}
\end{figure*}

\begin{figure*}[ht]
   \centering
   \begin{tabular}[b]{ccccccc}
   \includegraphics[width=0.24\columnwidth]{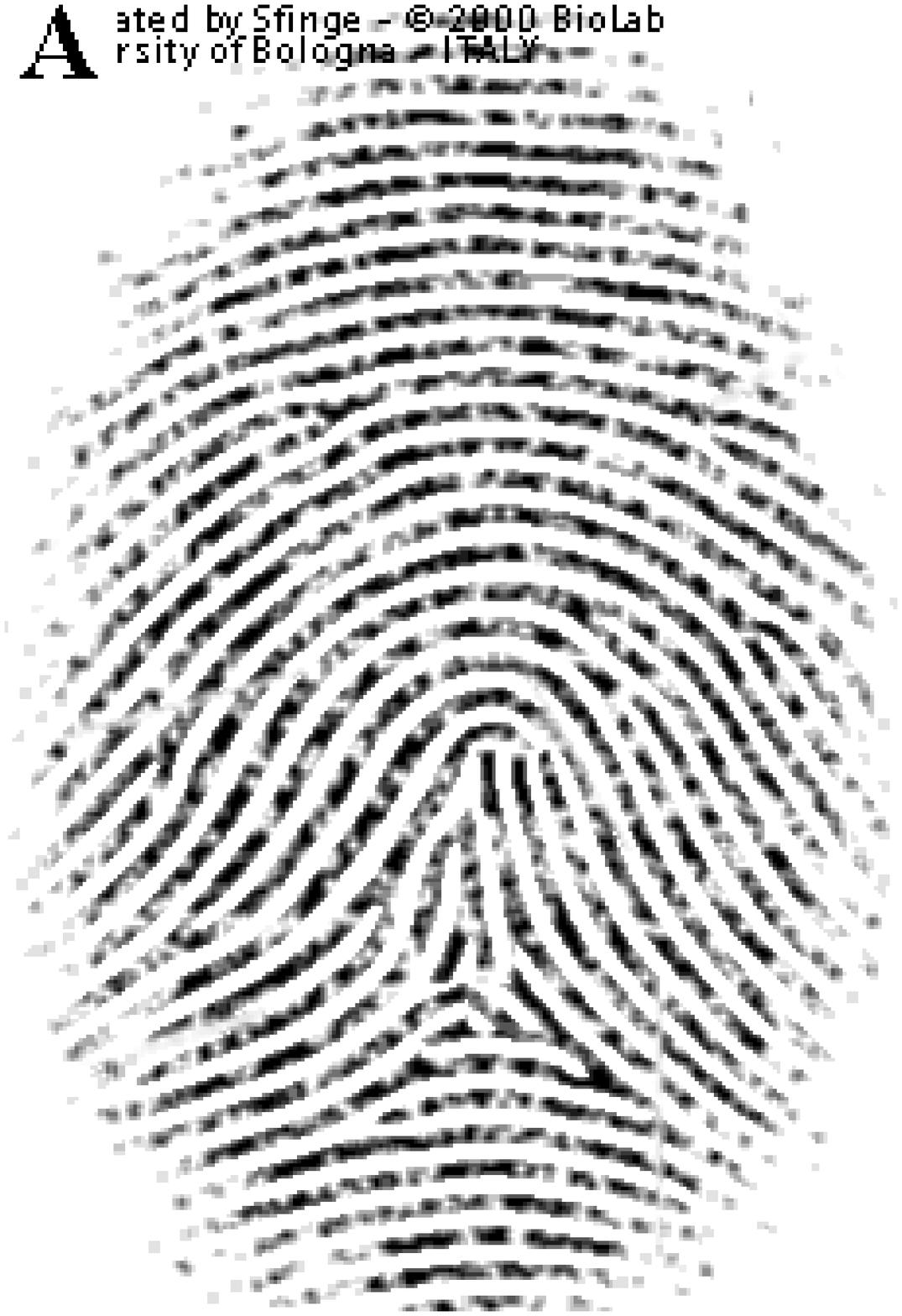} &
   \includegraphics[width=0.24\columnwidth]
   {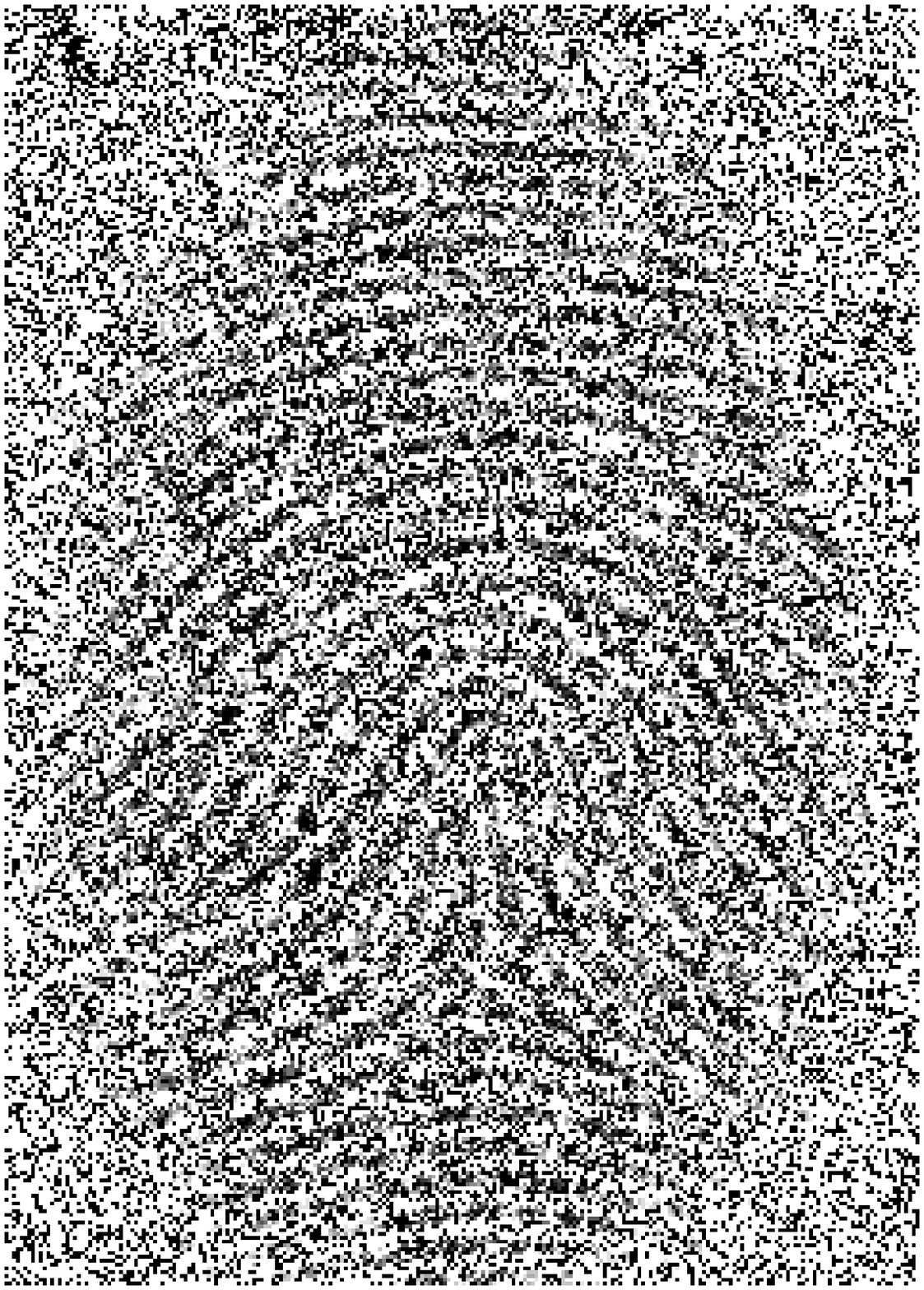} &
   \includegraphics[width=0.24\columnwidth]
   {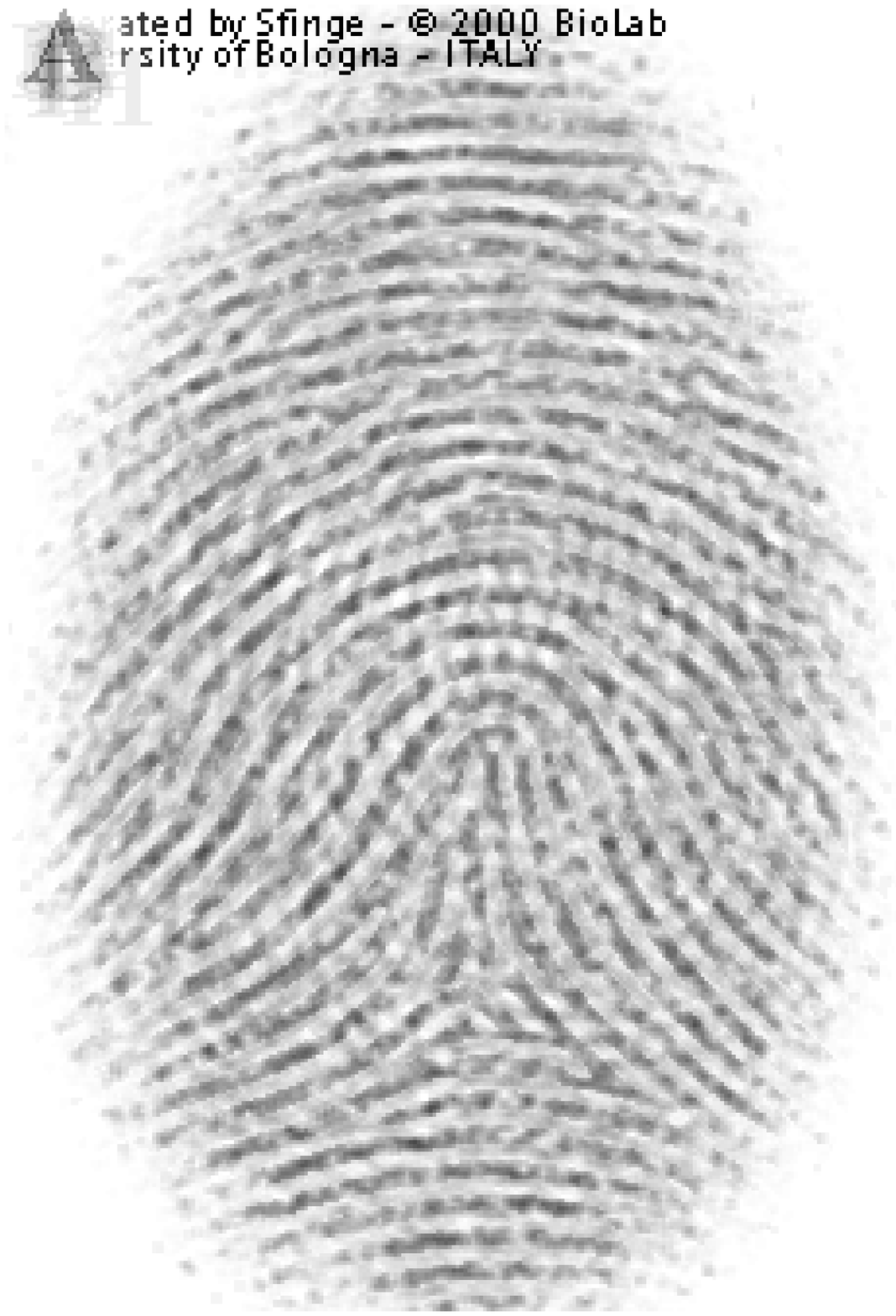} &
   \includegraphics[width=0.24\columnwidth]
   {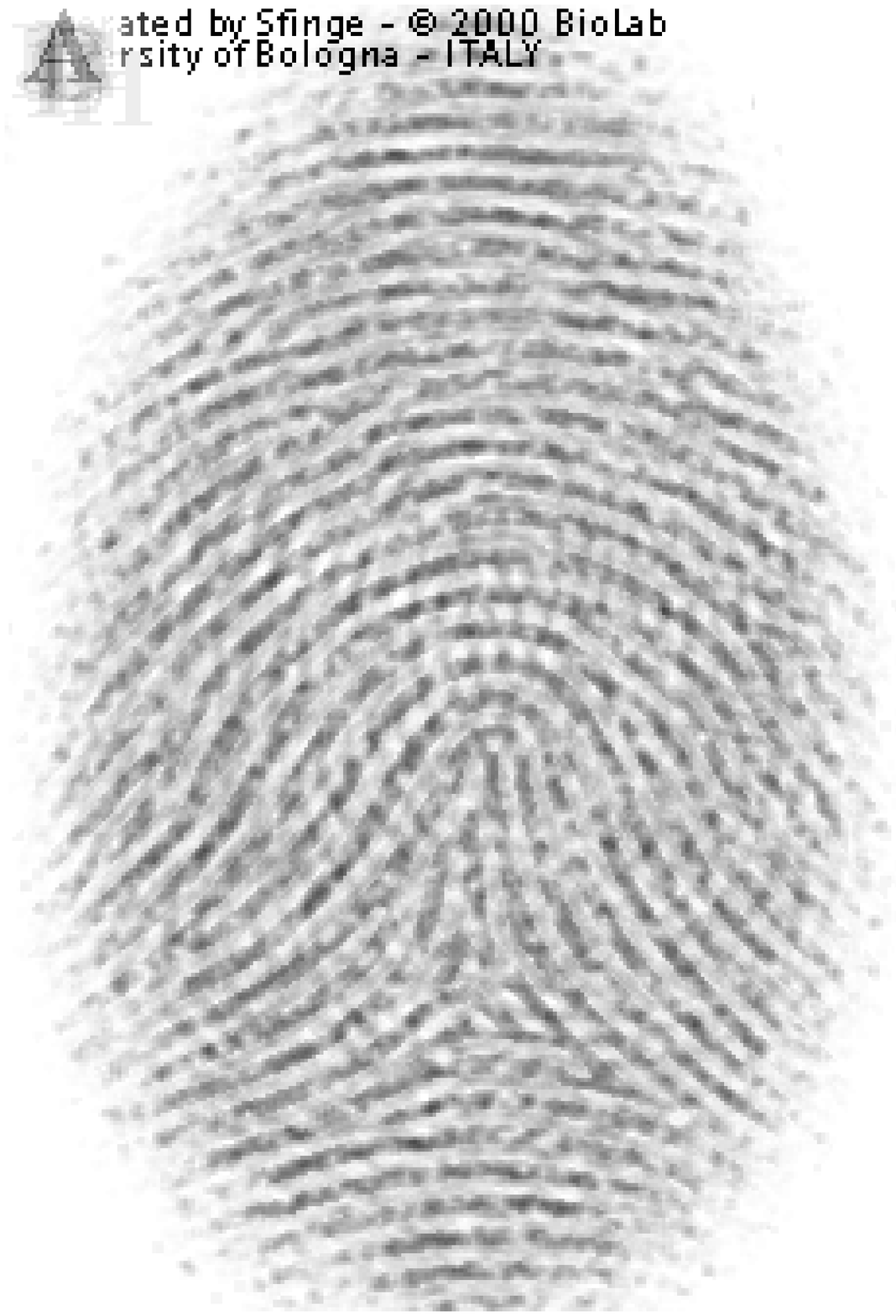} &
   \includegraphics[width=0.24\columnwidth]
   {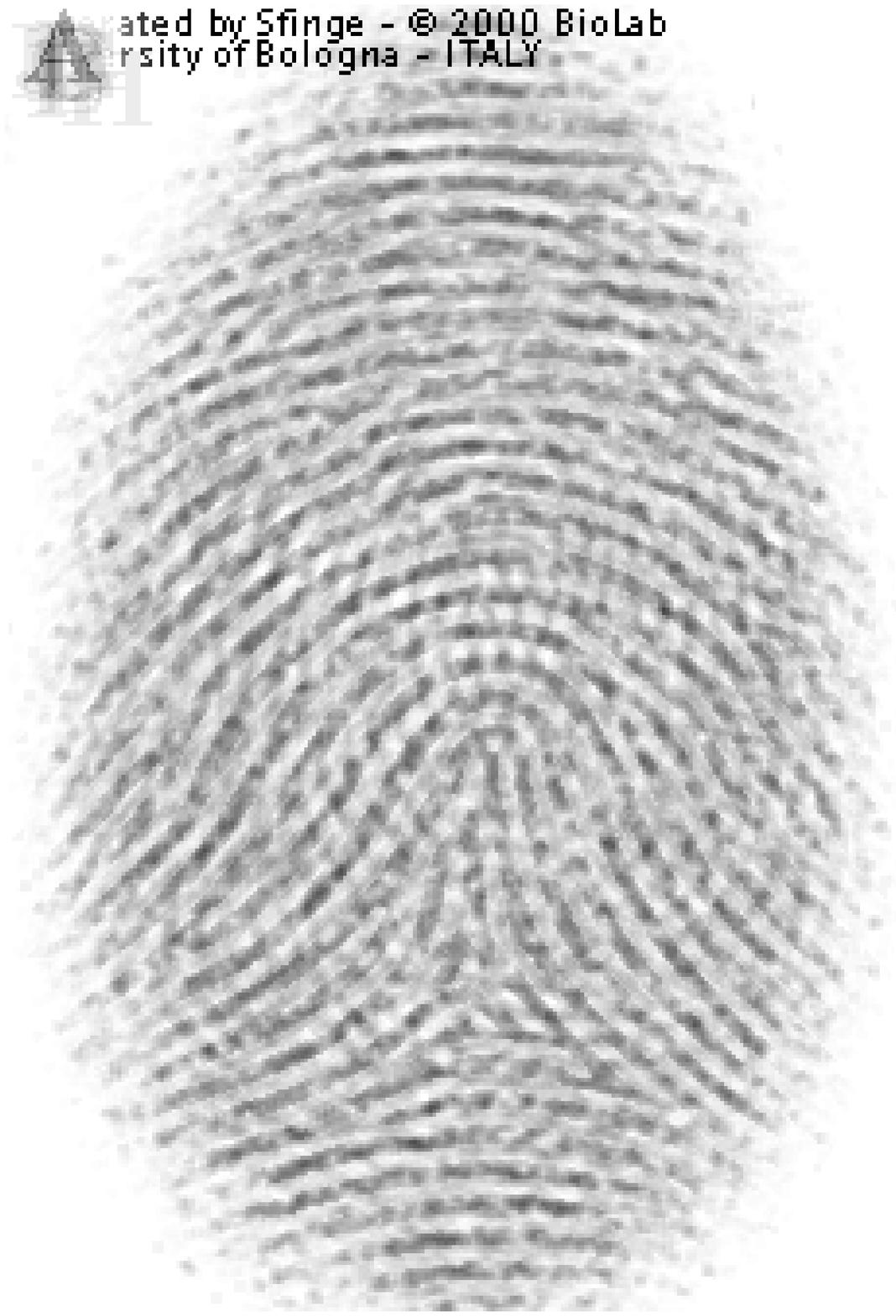} &
   \includegraphics[width=0.24\columnwidth]
   {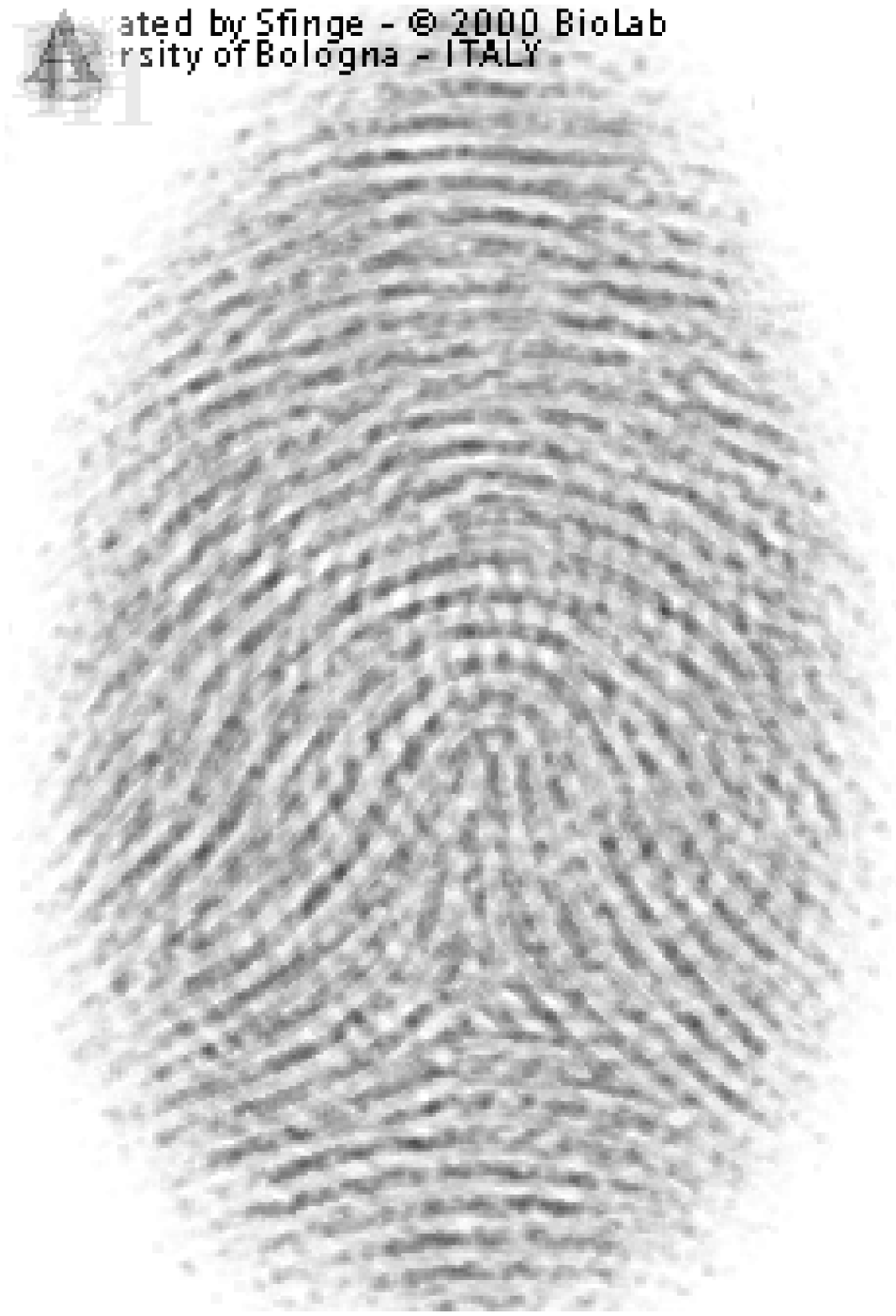} &
   \includegraphics[width=0.24\columnwidth]
   {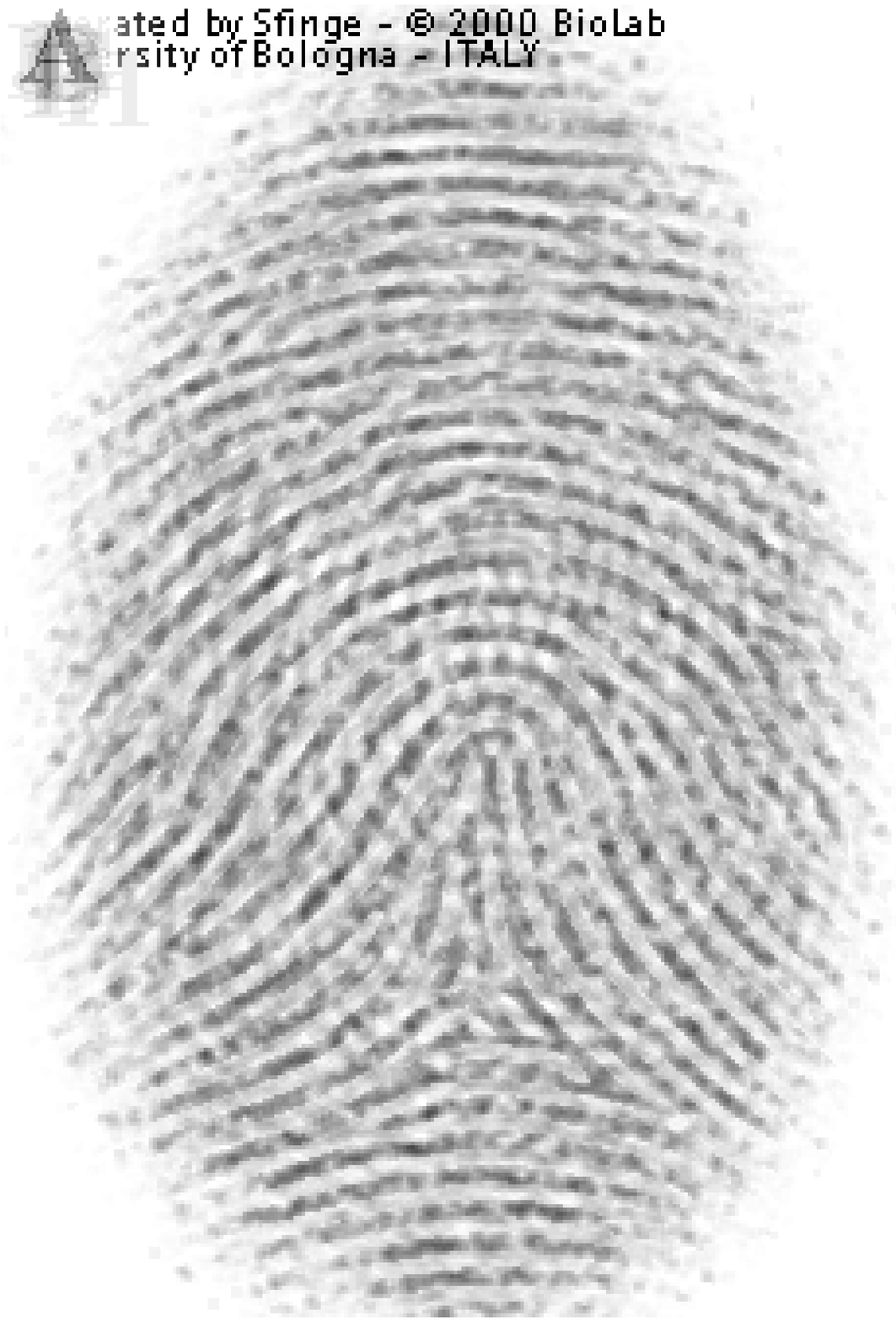} \\
   (a) & (b) & (c) & (d) & (e) & (f) & (g) \\
     \end{tabular}
   \caption{{\bf (a)} An original image.
   {\bf (b)} Original image (a) with 60\% salt-and-pepper noise. \
   {\bf (c)-(g)} Image restored from memory of 10 different
  simultaneously-stored fingerprints after presentation of the
  "query-image" which is:
  {\bf (c)} whole original image (a); \ {\bf (d)} 25\%-occluded image (a); \ {\bf (e)} 50\%-occluded (a); \
  {\bf (f)} 75\%-occluded (a); \
  {\bf (g)} noisy image (b).}
  \label{fingers}
\end{figure*}

\begin{figure*}[ht]
   \centering
   \begin{tabular}[t]{cccccc}
   \includegraphics[width=0.3\columnwidth]{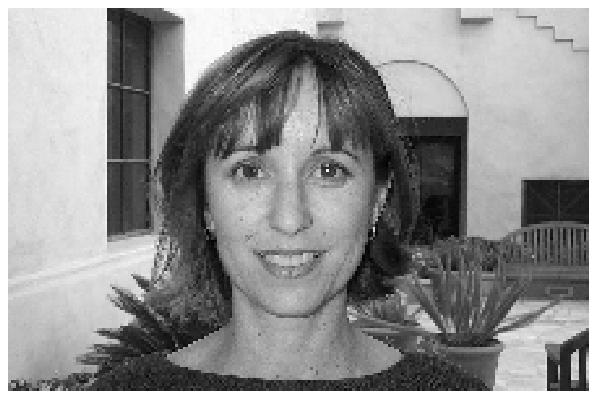} &
   \includegraphics[width=0.3\columnwidth]{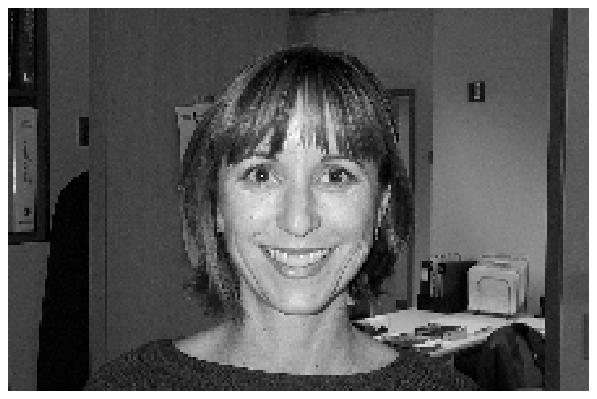} &
   \includegraphics[width=0.3\columnwidth]{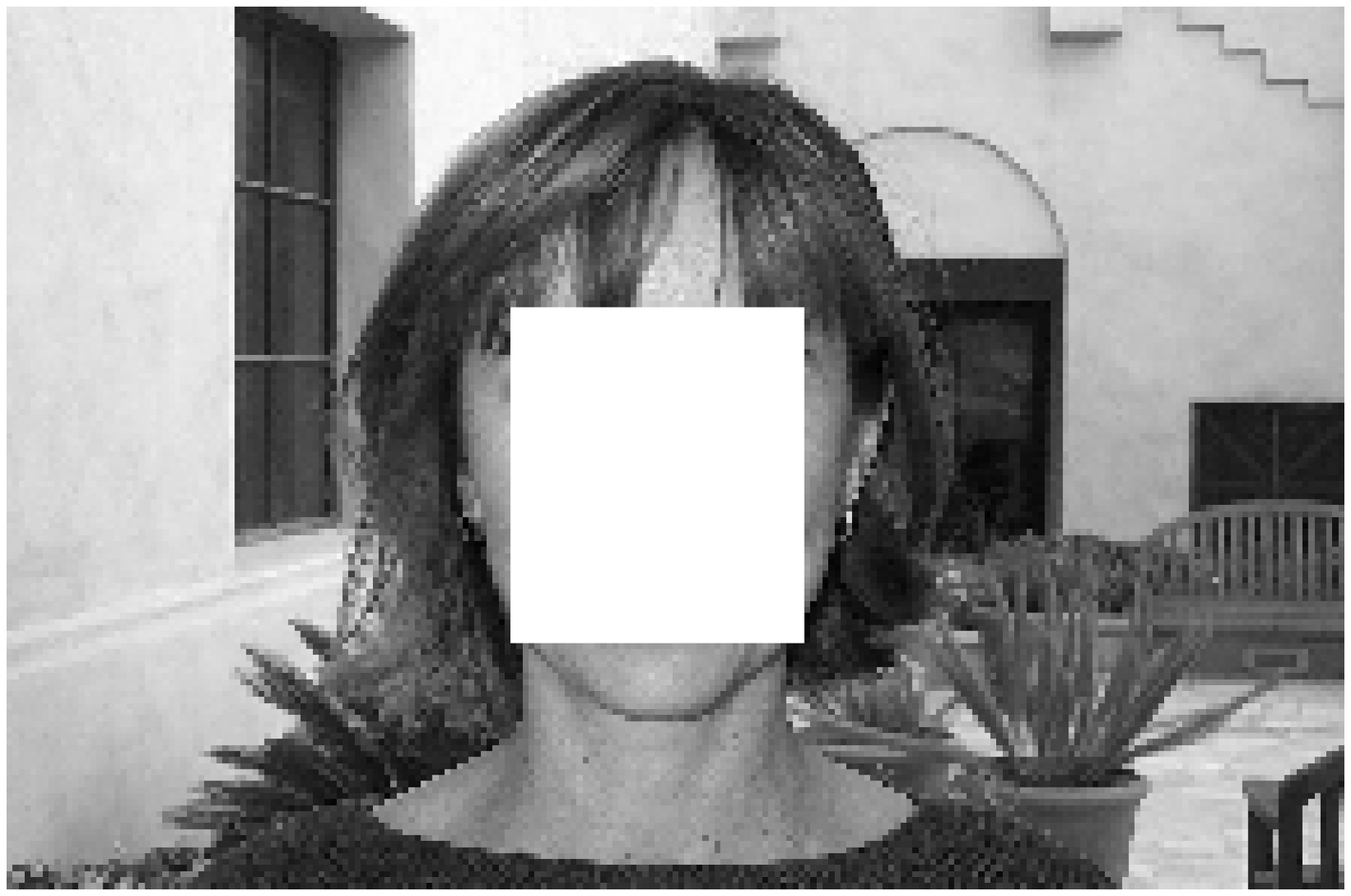} &
   \includegraphics[width=0.3\columnwidth]{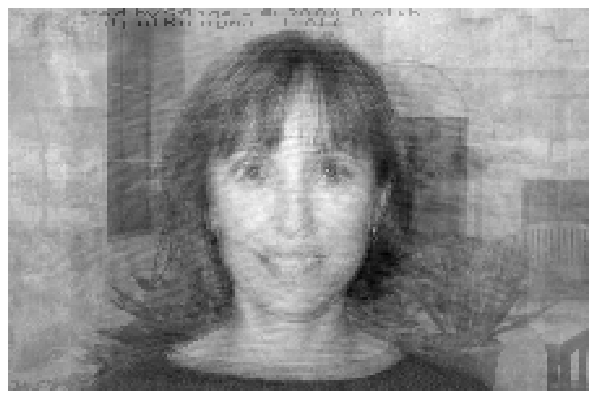} &
   \includegraphics[width=0.3\columnwidth]{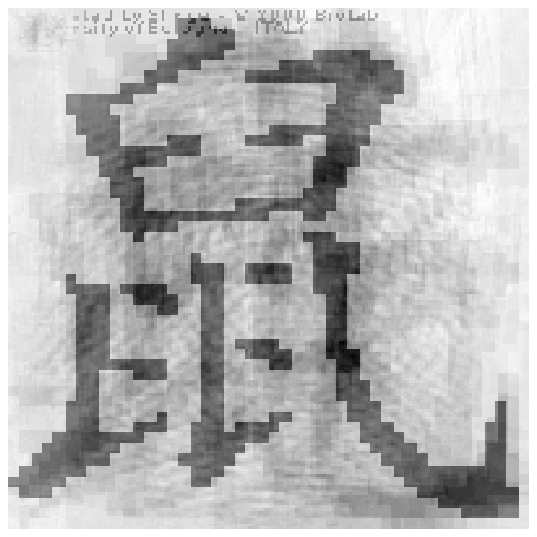} &
   \includegraphics[width=0.3\columnwidth]{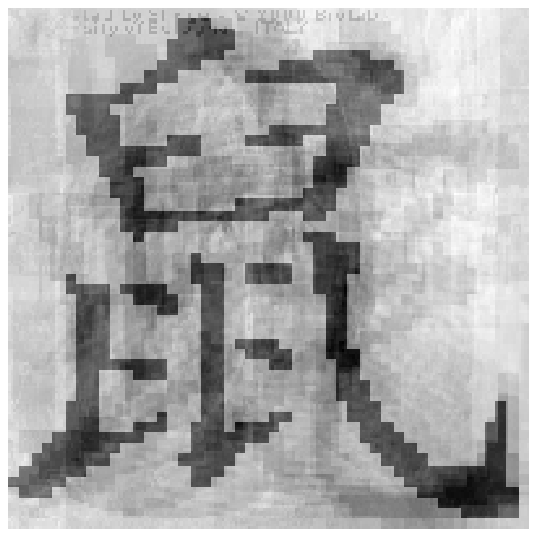} \\
   (a) & (b) & (c) & (d) & (e) & (f)  \\
     \end{tabular}
   \caption{Reconstruction from 30 simultaneously-stored images
   (10 different Chinese pictograms {\it and} 10 different
   fingerprints as on Fig. 2(a) {\it and} 10 different face-poses like on {\bf
   (a)\&(b)}: \ "Query" {\bf (c)} triggers reconstruction {\bf (d)}. \
   {\bf (e)\&(f)} Reconstructions from 25\%- and 50\%-occluded
   "query"-pictogram.}
  \label{mixed-set}
\end{figure*}

\section{Computational experiments}

The purpose of these experiments is just to verify the theory on
those basic aspects of the real quantum processing which {\it can}
be simulated. Real quantum-physical systems provide performance
(much) beyond what has been simulated by us up to now, and beyond
what is simulable at all.

All experiments were done on a Pentium-4 1.3-GHz PC using the
following algorithm programmed in MatLab with Image Processing
Toolbox:
\begin{itemize}
\setlength{\topsep}{0mm}
\setlength{\parskip}{0mm}\setlength{\partopsep}{0mm}
\setlength{\parsep}{0mm} \setlength{\itemsep}{0mm}

\item $P$ images with index $k$ were encoded into $
\vec{\tilde{v}}^k = ( \tilde{v}^k_1, ..., \tilde{v}^k_N ) $ where
a pixel's greyness is described by $\tilde{v}^k_j \in [0,255]$
($j=1,...,N$). \item Images were preprocessed according to: $
v^k_j = \tilde{v}^k_j - \frac{1}{N} \sum_{j=1}^N \tilde{v}^k_j $
for each $k, j$. The resulting vector $\vec{v}^k$ was then
normalized to satisfy $ \sum_{j=1}^N (v_j^k)^2 = 1 $. Such
normalized $\vec{v}^k$ are assumed to be quantum-implemented into
plane-wave/laser-beam $\psi^k$. \item Memory matrix, Eq. (1), was
calculated ("{\it storage} stage"). \item Later, in the "{\it
selective reconstruction} stage", a new "query / recall-key" input
was inserted. The network reacted as described in Eq. (3), or
equivalently in Eq. (6). The "query-input" was completed (if
partial initially) or corrected (if corrupted) based on memorized
examples, and scaled back into [0,255]-range.
\end{itemize}

Quality of reconstructed image $\vec{v}$ was measured with {\it
Peak Signal-to-Noise Ratio} (in dB; for 255 grey-values): \\
$ PSNR = 20 \log_{10} \left( \frac{255}{RMSE} \right) $ ;\\ $ RMSE
= \sqrt{\frac{1}{N} \sum_{j=1}^{N} ( v_j^{original} -
v_j^{reconstructed} )^2} $.

We found that {\it reconstruction-quality only slightly
decre\-ased with increasing number of images stored
simultaneously}, and that this behavior was {\it similar}
regardless of the type of stored images and the type and rate of
deviation of the query-image from the stored images. For two
examples see Fig. \ref{plots}. Compare these plots with Fig.
\ref{fingers} which de\-monstrates examples of "image recovery"
from occlusion or noise. Indeed, the capability of selective
reconstruction using memory is {\it almost the same for different
rates of degradation} (occlusion or corruption with noise) of the
query-image or its deviation from the original stored image(s).

The performance is indeed holography-like --- a small part of a
hologram contains enough information about the whole pattern
(stored in the hologram, our Eq. (5), in a parallel-distributed
way) that the whole can be retrieved from the small part.

As evident from Fig. \ref{fingers}, the image which shared most
pixels with the query-image, was selected from memory-matrix and
reconstructed ("recognized"), being disrupted (merely) by
cross-talk due to non-orthogonality of stored images. Such
results, typical for associative nets and holography, were got
also in the "mixed-set experiment" (Fig. \ref{mixed-set}). Here, 3
very-different sets of 10 different-content images, i.e. with big
inter-set differences and small intra-set differences, were
simultaneously stored. Cross-talk backgrounds can be seen in Figs.
\ref{mixed-set}(d-f), but the reconstructed images are not
disrupted too much.

\section{Conclusions}

Our simulations confirm Hopfield-net's capabilities. The novelty
of our simulations is reconsideration of Hopfield-net's
characteristics in the age of powerful computers -- early
simulations of the eighties had a limited resolution of patterns
rather than images. Moreover, our original proposal of
quantum-wave implementation opens a possibility of nets having up
to {\it almost infinite} size, and of processing of huge /
high-resolution images. Therefore, Hopfield-net's cross-talk and
storage limitations do not manifest (too) much for our practical
needs. The first problem, cross-talk, is reduced since {\it images
with a huge number of pixels are usually almost orthogonal}. The
second problem, memory-capacity of the Hopfield model is limited
(to $ P \doteq 0.14 N $), is much reduced with possibility of
"astronomically big" $N$. Since databases include limited-size
images, we have not yet been able to demonstrate the benefits of
(quantum) huge-image processing, but they {\it are} evident even
from classical holography \cite{opr,hol}.

Instead of plane-waves, images could be encoded into Gabor
wavelets \cite{wavelet} which are similar to quantum wave-packets.
Other possible (great) improvements will be studied in the future.

Our proposal is {\it enormously superior} to other proposed
quantum associative memories
\cite{ventura,ventopt,trugenberger,sch}, based on the mainstream
of the quantum computing science using quantum-implemented logic
gates, in the sense of simplicity, miniaturization, {\it natural}
physical realizability of associative processing, memory capacity
and dimensionality of data (specifically, size and resolution of
images). Models \cite{ventura,ventopt,trugenberger,sch} are,
however, more compatible with the mainstream attempts for an {\it
universal-purpose} quantum computer, not merely for associative
tasks which our model masters. \\

\footnotesize {\bf ACKNOWLEDGEMENTS}: \ \ M.P. thanks for
discussions to Prof. H.J. Caulfield, Prof. V. Bu\v zek, Drs. C.
Trugenberger and A. Vlasov. M.P. gratefully worked as EU
Marie-Curie post-doc fellow (contract no. HPMF-CT-2002-01808).


\end{document}